\documentclass{ptephy}
\usepackage{graphicx}
\usepackage{url}
\setcitestyle{numbers}
\begin{document}
\title{Amplitude dependent orbital period in alternating gradient accelerators}
\author{%
  \name{S. Machida}{1,\ast},
  \name{D. J. Kelliher}{1},
  \name{C. S. Edmonds}{2,6},
  \name{I. W. Kirkman}{2,6},
  \name{J. S. Berg}{3,}\thanks{This manuscript has been authored by employees of Brookhaven Science Associates, LLC under Contract Nos. DE-AC02-98CH10886 and DE-SC0012704 with the U.S.\ Department of Energy. The United States Government retains a non-exclusive, paid-up, irrevocable, world-wide license to publish or reproduce the published form of this manuscript, or allow others to do so, for United States Government purposes.},\\
  \name{J. K. Jones}{4,6},
  \name{B. D. Muratori}{4,6},
  \name{J. M. Garland}{5,6}
}
\address{
  \affil{1}{ASTeC, STFC Rutherford Appleton Laboratory, Harwell Oxford, Didcot, OX11 0QX, United Kingdom}
  \affil{2}{University of Liverpool, Liverpool, L69 7ZE, United Kingdom}
  \affil{3}{Brookhaven National Laboratory; Building 901A; PO Box 5000; Upton, NY 11973-5000; USA}
  \affil{4}{ASTeC, STFC Daresbury Laboratory, Daresbury, Warrington, WA4 4AD, United Kingdom}
  \affil{5}{University of Manchester, Manchester, M13 9PL, United Kingdom}
  \affil{6}{Cockcroft Institute of Accelerator Science and Technology, Daresbury, Warrington, WA4 4AD, United Kingdom}
  \email{shinji.machida@stfc.ac.uk}
}
\Year{2015}
%\pacs{29.20.-c, 29.27.-a, 41.85.-p}
\begin{abstract}%
  Orbital period in a ring accelerator and time of flight in a linear accelerator depend on the amplitude of betatron oscillations. The variation is negligible in ordinary particle accelerators with relatively small beam emittance. In an accelerator for large emittance beams like muons and unstable nuclei, however, this effect cannot be ignored. We measured orbital period in a linear non-scaling fixed-field alternating-gradient (FFAG) accelerator, which is a candidate for muon acceleration, and compared with the theoretical prediction. The good agreement between them gives important ground for the design of particle accelerators for a new generation of particle and nuclear physics experiments.
\end{abstract}
\maketitle
\thispagestyle{headings}
\section{Introduction}
A linear non-scaling Fixed-Field Alternating-Gradient (FFAG) accelerator~\cite{Symon1956,Kolomensky1966,Johnstone1999} was proposed to accelerate muon beams for a neutrino factory~\cite{Machida2003,Apollonio2009}. The fixed field nature makes very rapid acceleration possible so that short-lived particles such as muons can be accelerated. FFAG accelerators also have large acceptance, which is another essential requirement for acceleration of muons whose normalized emittance is a few 10,000 $\pi$ mm mrad even after ionization cooling. In addition, a proposed experiment for the search of charged lepton flavor violation called PRISM (Phase Rotated Intense Slow Muon source) utilizes a FFAG ring to convert a short bunch of muons with large momentum spread to a monochromatic long bunch~\cite{Kuno2000}. Although the FFAG of PRISM is a storage ring rather than an accelerator, large acceptance is one of the key ingredients of that experiment.

This paper describes the measurement of the time of flight in a linear non-scaling FFAG as a function of transverse amplitude. That the time of flight depends on transverse amplitude, and to lowest order the time of flight dependence is quadratic in the transverse phase space variables, is an obvious result of orbit geometry for a straight beamline. One of the first places it was identified as an important effect was in particle sources~\cite{Helm1968}, due to the large angular acceptances possible in a source that captures with solenoid focusing (see~\cite{Sayed2014} for the importance of the same effect in a high-intensity muon source). When using a small momentum compaction to get short bunches in an electron storage ring, this effect can be significant with respect to the energy-dependent time of flight, and thus can have an impact on longitudinal dynamics~\cite{Bruck1973,Deacon1980,Deacon1981,Emery1993}. In free electron lasers, this effect can cause particles with higher transverse amplitudes to lose synchronism with the coherent bunch structure~\cite{Deacon1980,Deacon1981,Artamonov1989}. Ionization cooling channels for muons also require very large angular acceptances; the dependence of the time of flight on transverse amplitude requires an introduction of a correlation between beam energy and transverse amplitude to avoid unwanted longitudinal emittance growth~\cite{Lebrun1998,Fernow1999,Ankenbrandt1999}. Finally, and of particular interest for this work, this phenomenon is important for non-scaling FFAGs, which typically operate on both sides of the time of flight minimum, and for some important applications, particularly muon acceleration, accelerate beams with large transverse emittances~\cite{Machida2006,Berg2007}.

To achieve extremely large transverse acceptances, a linear non-scaling FFAG relies on the exclusive use of linear magnets. The chromaticity of the machine is thus the uncorrected natural chromaticity. Introduction of sextupoles to correct the chromaticity results in a significant reduction in the transverse acceptance~\cite{Berg2010}.

There is a simple relationship between this time of flight dependence on the transverse amplitude and the chromaticity, given by
\begin{equation}
  \Delta t = -\dfrac{2\pi}{c}(\xi_hJ_h+\xi_vJ_v),
  \label{eq:1}
\end{equation}
where $c$ is the speed of light, $\xi_{h,v}$ is chromaticity ($p(d\nu_{h,v}/dp)$, with $\nu_{h,v}$ the tune per period and $p$ the total momentum) in horizontal and vertical direction, and $J_{h,v}$ is action variable of betatron oscillations (defined as usual such that the maximum position a particle with action $J_{h,v}$ could have would be $\sqrt{2\beta_{h,v}J_{h,v}}$, with $\beta_{h,v}$ being the Courant-Snyder beta function). We assume that particles are ultra-relativistic. The period (turn, superperiod, cell, etc.) used for computing $\Delta t$ and $\xi_{h,v}$ should correspond. The expression follows immediately from describing the system via a Hamiltonian. The earliest derivation of this relationship that we are aware of is in~\cite{Artamonov1989}; other authors (\cite{Forest1998}, p.~140, and \cite{Berg2007}) have reproduced a similar result from a Hamiltonian. Others (the most well-known example being~\cite{Emery1993}) have derived similar results without directly using Hamiltonian techniques, but as pointed out in~\cite{Forest1998}, pp.~264--265, this has often led to incorrect assumptions about the chromaticity factors in~(\ref{eq:1}): the chromaticities in that equation arise from the derivatives of the tunes with energy, with no qualifiers as to the cause of the energy variation.

The physical mechanism by which correcting chromaticity via sextupoles reduces the time of flight variation with transverse amplitude is not immediately obvious. A sextupole provides a vertical magnetic field that varies quadratically with both horizontal and vertical position. While a particle with no betatron amplitude sees no field, a particle with a nonzero betatron amplitude sees a nonzero average field from that sextupole when averaged over several turns, and that average field is proportional to the particle's $J_{h,v}$ (quadratic in the transverse position/momentum). The sextupole therefore acts like a dipole whose field is proportional to $J_{h,v}$, and therefore results in a closed orbit displacement and change in orbit length proportional to $J_{h,v}$. This effect is discussed in somewhat more detail in Sec.~2 of~\cite{Berg2007}, and a similar mechanism in a helical wiggler is discussed in the second to last paragraph of~\cite{Artamonov1989}.

A linear non-scaling FFAG called EMMA (Electron Model for Many Applications) was built and commissioned at Daresbury Laboratory~\cite{Barlow2010,Machida2012}. It gave the unique opportunity for a direct measurement the amplitude dependent orbital period, that is most readily possible with large acceptance machines like EMMA, although the difference in orbital period is still very small, of the order of ps, and measuring it is extremely difficult. We first reported our measurement in~\cite{Muratori2013}, though only the result was briefly mentioned; in this paper we describe the measurement and analysis in more detail. Our measurement is unique and of interest because it directly measures, to the extent possible, all of the quantities in Eq.~(\ref{eq:1}): we have a direct measurement of the orbital period from a BPM signal averaged over multiple turns; we directly measure the betatron oscillation amplitude by measuring phase space coordinates via a pair of BPMs separated only by a drift and finding the beam ellipse these phase space coordinates trace out; and finally we have tunes measured at several energies by determining the beam's oscillation frequency, from which we compute chromaticity as the derivative of a curve fit through those tunes. An experiment reported by Shoji \textit{et al.}~\cite{Shoji2014} also demonstrates the results of Eq.~(\ref{eq:1}), but in that experiment, some parameters in Eq.~(\ref{eq:1}) are obtained by measuring different quantities and connecting them to quantities in the equation via known machine parameters.

The measurement result agrees with the expected relationship in Eq.~(\ref{eq:1}). In the following, we will first explain the experimental setup and measurements, then we will give the results of our measurements and show how well they fit the expected relationship.

\section{Experimental Setup}
EMMA is an accelerator to demonstrate and study in detail the properties of a linear non-scaling FFAG. Its lattice consists of 42 identical doublet cells with focusing and defocusing quadrupole magnets. A beam traverses the quadrupoles off-axis by design and obtains a net bending. It was designed to accelerate a single bunch of 10.5~MeV/$c$ electrons to 20.5~MeV/$c$ through the serpentine channel~\cite{Berg2001,Johnstone2003}. The circumference is about 16.5~m and the orbital period is about 55.3~ns. Although the horizontal orbit does not stay constant during acceleration because of the fixed field lattice magnets, the dispersion function had been minimized and the normalized physical acceptance should be at least 3000~$\pi$\,mm\,mrad in both horizontal and vertical direction within a vacuum chamber whose diameter is about 40~mm.

\subsection{Measurement of Orbital Period}
\begin{figure}
  \begin{center}
    \includegraphics[width=0.5\linewidth]{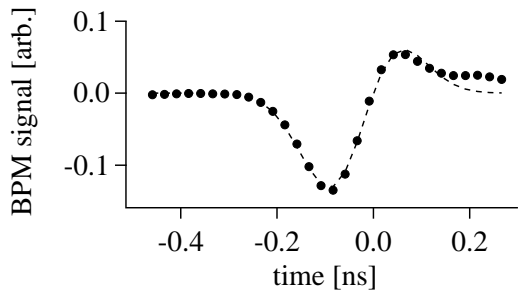}
  \end{center}
  \caption{The raw BPM signal (dots) and a function of the form $f(x)=(c_0+c_1x)\exp[-x^2/(2\sigma^2)]$ fit to that signal (dashed line).}
  \label{fig:1}
\end{figure}
The orbital period is about 55.3~ns, while the shift due to amplitude dependence is the order of ps. The arrival time of the bunch for each pass was obtained by fitting a measured beam position monitor (BPM) signal, which is a filtered differential of the bunch current, recorded on a 40~GS/s oscilloscope, to a function of the form $f(x)=(c_0+c_1x)\exp[-x^2/(2\sigma^2)]$ and determining the zero crossing, which was used as the arrival time, as shown in Fig.~\ref{fig:1}. The orbital period was then obtained to high precision by averaging over a number of turns. When we began studying EMMA, we used the internal clock of the oscilloscope as a reference time, but this proved problematic since the oscilloscope appeared to occasionally drop a sample, resulting in orbital period fluctuations corresponding to the oscilloscope sample period. Instead, we simultaneously recorded a high precision 1.3~GHz reference waveform used for timing the entire facility, and performed our timing measurements by finding the phase of the BPM signal with respect to that reference waveform.

\subsection{Measurement of Tunes and Computation of Chromaticity}
All the momentum dependent parameters in EMMA were measured not by changing the beam momentum, but by equivalently changing the excitation current of the lattice magnets, while keeping the beam momentum fixed at 12.5~MeV/$c$. The dynamics of a beam with a momentum of $p$ with magnet currents at their reference values for 12.5~MeV/$c$ are identical to the dynamics of a 12.5~MeV/$c$ beam with the magnets' currents scaled by a factor of 12.5/$p$, with the exception of a change in the particle velocity. We refer to this lattice as having an ``equivalent momentum'' of $p$~MeV/$c$.

\begin{table}[!tbp]
  \caption{\label{tab:4}Measured vertical cell tunes and their uncertainties computed using a discrete Fourier transform of 10 turns of data. Tune values are the peak in the discrete Fourier transform, the uncertainties arise from the observed width of the peak, which for most cases arises from the spacing of frequency values in the discrete Fourier transform.}
  \begin{center}
    \begin{tabular}{lrrrr}
      \hline
      Equivalent&Horizontal&Horizontal&Vertical&Vertical\\
      Momentum&Tune&Tune&Tune&Tune\\
      (MeV/$c$)&&Uncertainty&&Uncertainty\\
      \hline
      12&0.2506&0.0012&0.2196&0.0012\\
      14&0.2100&0.0012&0.1814&0.0012\\
      16&0.1814&0.0012&0.1504&0.0012\\
      18&0.1599&0.0012&0.1289&0.0012\\
      20.3&0.1480&0.0048&0.1026&0.0012\\
      \hline
    \end{tabular}
  \end{center}
\end{table}
Betatron oscillation around the ring was measured at BPMs in all the 42 cells for several equivalent momenta. Since all the cells are designed to be identical, we defined the cell tune to be the betatron phase advance through each cell divided by $2\pi$ (the tune for a full turn is obtained by multiplying this number by 42). We performed a discrete Fourier transform of the BPM data for 10 turns (420 cells) and chose the maximum amplitude from that to be the cell tune, the resulting values for which are given in Table~\ref{tab:4}. To determine the chromaticity per cell at the equivalent momentum $p_0= 17.6\text{ MeV}/c$, we first make a weighted least-squares fit of the cosines of $2\pi$ times the cell tunes to a polynomial in the inverse of the equivalent momentum. The chromaticity per cell at $p_0$ can then be determined evaluating the polynomial and its derivative at $1/p_0$, applying the appropriate derivative composition formulas to find the derivative of the tune with respect to momentum, then multiplying the result by $p_0$. Linear error propagation then gives the uncertainty in the chromaticity. The momenta, tunes, and the uncertainties in the vertical tunes are given in Table~\ref{tab:4}. Horizontally, the best fit is with a cubic polynomial, resulting in a chromaticity per cell of $-0.158\pm0.014$, with a chi-squared per degree of freedom of 0.7. Vertically, the best fit is with a quadratic polynomial, resulting in a chromaticity per cell of $-0.204\pm0.006$, with a chi-squared per degree of freedom of 1.5. To obtain chromaticities per turn, multiply the chromaticity per cell by 42.

\subsection{Measurement of Betatron Amplitude}
In order to set the betatron oscillation amplitude in the vertical direction, we varied the deflection angle at a vertical orbit corrector in the injection beam line. The transverse beam emittance from the pre-accelerator is small compared with the acceptance of EMMA so that the whole bunch oscillates coherently as a single particle until tune spread gradually damps the coherent signal.

\begin{figure}
  \begin{center}
    \includegraphics[width=0.5\linewidth]{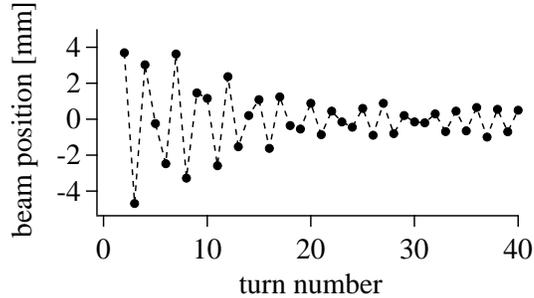}
  \end{center}
  \caption{Measured beam position signal over a number of turns, showing the reduction of signal amplitude due to tune spread arising from chromaticity and energy spread in the beam.}
  \label{fig:2}
\end{figure}
This tune spread arises because EMMA was designed to eliminate nonlinearity as much as possible to ensure a large dynamic aperture. The natural chromaticity remains and induces tune spread as a result of the finite momentum spread. For example, a momentum spread of 0.2\% leads to a tune spread of roughly $2\times10^{-2}$ per turn. A point-like beam injected off axis in the transverse phase space ends up as a ring in phase space within 25 turns. Since the BPM detects only the charge center of particles, measured betatron oscillations appear to damp in the same time scale, as can be seen in Fig.~\ref{fig:2}.

We measured beam positions at monitors located at both ends of three different straight sections in EMMA. The beam position monitors were far enough from the adjacent quadrupole magnets that the intervening straight could be treated as a drift; thus we could infer position and angle of the beam orbit in phase space at the center of the drift from the beam position data at the two ends of each straight. Were it not for the decoherence of the BPM signal due to energy spread in the beam and the nonzero chromaticity, these data could be used to directly reconstruct the action value.

To compute the action in the presence of the chromaticity-driven decoherence, a somewhat more involved procedure is required, which is outlined here and described in more detail in~\cite{Edmonds2014}. Begin with a complex representation of the position and momentum measurements as
\begin{equation}
  f_n = \dfrac{y_n}{\sqrt{\beta_y}}+i\left(p_{y,n}\sqrt{\beta_y}+\dfrac{\alpha_yy_n}{\sqrt{\beta_y}}\right),
  \label{eq:2}
\end{equation}
where $y_n$ and $p_{y,n}$ are respectively the transverse position and momentum of the bunch centroid at the observation point on the $n^{\text{th}}$ turn. Before computing $y_n$ and $p_{y,n}$ from the beam position monitor data, the average of each individual monitor's position data over several turns was first subtracted from that data to remove any closed orbit offset. $\beta_y$ and $\alpha_y$ are the Courant-Snyder parameters at the observation point. $\beta_y$ and $\alpha_y$ are not initially known, but one has an initial estimate for them from the lattice design. Next, construct the quantity $g_n=f_ne^{2\pi inQ_{h,v}}$, where $Q_{h,v}$ is the tune per turn whose measurement we describe above. In principle the complex magnitude of $g_n$ is decreasing due to the decoherence, but its complex phase, which is the initial phase of the betatron oscillation ($-\varphi_0$), should remain constant. We therefore choose values of $\beta_y$ and $\alpha_y$ that minimize the RMS variation with $n$ of the complex phase of $g_n$. The values of $\beta_y$ and $\alpha_y$ that are found during this optimization process are used later when calculating the momentum distribution of the bunch. 

Taking into account the energy spread in the beam and the chromaticity, $f_n$ can now be expressed as
\begin{equation}
  f_n\approx\sqrt{2J_{h,v}}e^{-i\varphi_0}\int_{-1/(2\xi_{h,v})}^{1/(2\xi_{h,v})}e^{-i2\pi n(Q_{h,v}+\xi_{h,v}\delta)}\Phi(\delta)\,d\delta
  \label{eq:3}
\end{equation}
where $\Phi(\delta)$ is the momentum distribution function, $\delta$ is the momentum deviation from the average, and $\xi_{h,v}$ is the chromaticity per turn, whose measurement is discussed in the next subsection. In Eq.~(\ref{eq:3}), it is assumed that the contribution to the tune spread of the bunch due to both the nonlinear terms of the chromaticity and the emittance of the bunch coupled with amplitude detuning is negligible. We then apply a discrete Fourier transform to Eq.~(\ref{eq:3}); since $\Phi(\delta)$ is a distribution function normalized to 1, we can ignore multiplicative constants and compute
\begin{equation}
  \Phi(\delta)\propto e^{i\varphi_0}\sum_{n=-N}^{N}f_ne^{i2\pi n(Q_{h,v}+\xi_{h,v}\delta)}
  \label{eq:4}
\end{equation}
where $N$ is the total number of turns for which there is measured data, and the terms for the negative turn numbers may be found by considering the complex conjugate of $f_n$,
\begin{equation}
  f_{-n}=e^{-i2\varphi_0}f^*_n.
  \label{eq:5}
\end{equation}
Now that $\Phi(\delta)$ is known, the only remaining unknown is $J_{h,v}$, which is found by a fit of Eq.~(\ref{eq:3}) to the BPM data.

\section{Results and Summary}
\begin{figure}
  \begin{center}
    \includegraphics[width=0.75\linewidth]{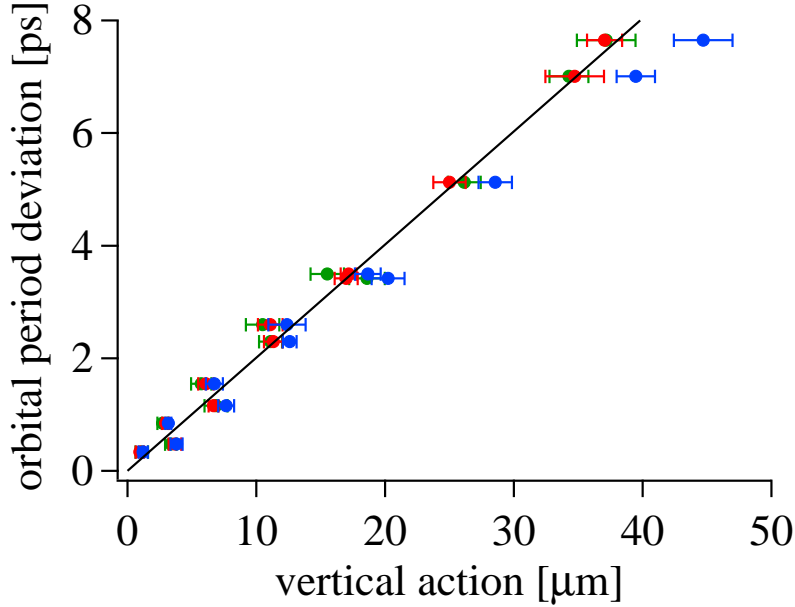}
  \end{center}
  \caption{Points are measured values of the increase in orbital period and vertical action. Different colors represent different BPM locations. The line is a linear fit to the data.}
  \label{fig:3}
\end{figure}
Figure~\ref{fig:3} shows the measured orbital period as a function of vertical betatron oscillation amplitudes at the equivalent momentum of 17.6~MeV/$c$. In performing the measurement, we first set our injection magnets to minimize the betatron oscillation amplitude in the ring for both the horizontal and vertical planes. Then we applied 15 different settings for the vertical corrector in the injection line to create vertical betatron oscillations in the ring of varying amplitudes. The corrector currents were applied in a random order to reduce any effects of secular drifts in machine parameters, such as beam energy. 

For each vertical corrector setting, we performed 50 measurements of the vertical action $J_v$, using the average for the result and the standard deviation in the average for the uncertainty. For each setting of the vertical orbit corrector, the initial amplitude was measured at three locations in the ring. Corrector settings resulting in the smallest amplitudes are omitted because we could not obtain accurate action values for them.

The time of flight for each setting was obtained by averaging the time of flight for the first 50 turns, then taking the average over 20 injections. The uncertainty is computed as the standard deviation of the latter average, and is at or below 0.1~ps for every setting.

To assess the agreement between our data and the predicted chromaticity, we perform a linear fit to the orbital period and action data. We first assign a single action value and uncertainty for each vertical corrector setting, since the measurements at the three different ring positions cannot be treated as uncorrelated for the purposes of the linear fit. We then perform the linear fit to the resulting points, but must address larger-than-expected fluctuations in the data as described below.

To compute the value for the action for a particular corrector setting, we first compute the uncertainty-weighted average. For some settings, the chi-squared per degree of freedom in that average is large, indicating systematic effects that were not accounted for. To avoid overweighting settings where that was the case, the uncertainty assigned is a weighted average of the uncertainty computed through error propagation and the uncertainty from the weighted average, with the weighting factor being the chi-squared distribution function. 

When we perform a weighted least-squares linear fit to the resulting points, the result of which is shown in Fig.~\ref{fig:3}, we find a chi-squared per degree of freedom of 9, indicating either unaccounted-for uncertainties or that the linear model is poor. Qualitatively, the figure indicates that the linear model is a good representation of the data. To account for the excessive uncertainties, we increase the underlying uncertainties by a factor of the square root of the chi-squared per degree of freedom. The resulting slope is $0.201\pm0.017$~ps/$\mu$m, which from Eq.~(\ref{eq:1}), would correspond to a chromaticity per cell of $-0.229\pm0.019$. This is consistent with the chromaticity per cell of $-0.204\pm0.006$ predicted from the tune measurements as discussed above.

In summary, for several betatron amplitudes, we have made direct measurements of the time of flight and the betatron amplitude. Using measured tune values to compute the chromaticity, we have shown that Eq.~(\ref{eq:1}) is consistent with our experimental results, though unaccounted-for measurement uncertainties complicate the analysis. This effect, the time of flight dependence on transverse amplitude, is of great importance~\cite{Machida2006,Berg2007} for proposed accelerators requiring large transverse amplitudes, such as muon accelerators~\cite{Lebrun1998,Fernow1999,Ankenbrandt1999,Machida2003,Apollonio2009,Sayed2014}, where chromaticities are generally large, as well as for PRISM~\cite{Kuno2000}, should its chromaticity be imperfectly corrected or a non-scaling FFAG option be considered. The large transverse acceptance of the non-scaling FFAG EMMA~\cite{Barlow2010,Machida2012} enabled this direct measurement to be made straightforwardly.

\section*{Acknowledgments}
The work is supported by the BASROC/CONFORM project under the EPSRC grant No. EP/E032869, the US Department of Energy under contract Nos. DE-AC02-98CH10886, DE-SC0012704, and DE-AC02-07CH11359, and STFC. We would like to thank the entire ALICE/EMMA team. We also would like to thank Drs. M. W. Poole, C. R. Prior, S. L. Smith and A. Wolski for their encouragements.
\bibliographystyle{ptephy}
\bibliography{1509-022-3G-ShinjiMachida-01}
\end{document}